\documentclass[aip,pop,amsfonts,amsmath,amssymb,table,preprint]{revtex4-1}
\pdfoutput=1 

\usepackage{graphicx}

\usepackage[english]{babel}
\usepackage[utf8]{inputenc}
\usepackage[pdftitle={Article}, pdfauthor={Author}]{hyperref} 

\begin{document}


\title{Numerical modeling of two magnetized counter-propagating weakly collisional plasma flows in arch configuration} 

\author{Artem V. Korzhimanov}
\email[Correspondence email address: ]{artem.korzhimanov@ipfran.ru}
\author{Sergey A. Koryagin}
\affiliation{Federal Research Center A. V. Gaponov--Grekhov Institute of Applied Physics of the Russian Academy of Sciences, Nizhny Novgorod, Russia}
\affiliation{Lobachevsky State University of Nizhny Novgorod, Nizhny Novgorod, Russia}
\author{Andrey D. Sladkov}
\affiliation{Federal Research Center A. V. Gaponov--Grekhov Institute of Applied Physics of the Russian Academy of Sciences, Nizhny Novgorod, Russia}
\author{Mikhail E. Viktorov}
\affiliation{Federal Research Center A. V. Gaponov--Grekhov Institute of Applied Physics of the Russian Academy of Sciences, Nizhny Novgorod, Russia}
\affiliation{Lobachevsky State University of Nizhny Novgorod, Nizhny Novgorod, Russia}

\date{\today}

\begin{abstract}
Numerical modeling of the interaction process of two counter-streaming supersonic plasma flows with an arched magnetic field configuration in the regime of a magnetic Mach number of the order of unity $M_m \sim 1$ is carried out. The flows were launched from the bases of the arch along the direction of the magnetic field. It is shown that the interaction has non-equilibrium and non-stationary nature. It is accompanied by an expansion of the resulting magnetic plasma arch due to $E\times B$ drift with the formation of a region with oppositely directed magnetic fields, in which magnetic reconnection is observed. In the subcritical regime $M_m < 1$ the reconnection process is slow, and in the overcritical one $M_m > 1$ it is more intense and leads to plasma turbulization. Filamentation of flows due to the development of Weibel instability, as well as excitation of surface waves near the ion-cyclotron frequency on the surface of the plasma tube are also observed. The modeling was carried out for the parameters of an experiment planned for the near future, which made it possible to formulate the conditions for observing the effects discovered in the modeling.
\end{abstract}


\maketitle 

\section{\label{sec:intro}Introduction}

In astrophysics, the interaction of plasma flows with magnetic structures play a major role. In particular, in the atmospheres of stars and the magnetospheres of planets, magnetic field configurations in the form of an arch, usually filled with plasma, are widespread. Such magnetized plasma arch structures have been observed on the Sun \cite{Masuda1994} and in the Earth's magnetosphere \cite{Liu2020}. It has been shown that they play a significant role in solar flares \cite{Xia2020}, as well as in various explosive processes accompanied by the reconnection of magnetic lines \cite{Yamada2010}.

In addition to space observations, magnetized plasma arches are also studied in the laboratory \cite{Katz2010, Tripathi2010, Stenson2012}. In controlled conditions, it is possible to conduct a more detailed analysis of the processes occurring in the system and give them a physical explanation. However, due to laboratory limitations, studies are conducted on objects much smaller than space ones and at much shorter times. In this case, the correspondence between laboratory modeling and space observations is achieved by comparing dimensionless parameters that allow scaling of physical processes. Nevertheless, not all interaction regimes observed in space can be implemented in the laboratory within the framework of one or two experimental setups, so it is of interest to develop new approaches to laboratory modeling. Recently, our group proposed and implemented an original method for creating a magnetized plasma arch during an arc discharge in a magnetic field created by two coils located at an angle to each other \cite{Viktorov2015}. The features of the resulting configuration include the supersonic nature of the plasma expansion, magnetic Mach numbers in a wide range $M_m\sim 0$--$10$, and the small rate of the collisions compared to the gyrofrequency. As a result, the resulting plasma is non-equilibrium and the interaction is accompanied by the development of instabilities, including those of a kinetic nature \cite{Viktorov2017}.

Despite the rich possibilities for performing various diagnostics in a laboratory experiment, many processes occurring in it have too small scales in time and space to be distinguished by modern methods. In this regard, numerical modeling is widely used to plan experiments and explain their results. As an example, we can cite a recent work on the numerical modeling of the interaction process of a plasma flow incident on a magnetic arch \cite{Wang2024} in a configuration similar to the experiment on our setup \cite{Viktorov2019a, Viktorov2019b}. At present, work is being carried out on the setup to study the collision process of two counter-streaming flows injected from the bases of a magnetic arch along magnetic lines. The present work is aimed at the numerical modeling of this process and the identification of the main interaction modes and the physical processes accompanying them. The main attention is paid to the emerging structures of plasma density, directly observed by its optical glow in the experiment, as well as to the processes of development of various instabilities and, in particular, the Weibel instability \cite{Weibel1959}.

\section{\label{sec:setup}Experimental setup}

\begin{figure}
\includegraphics[width=0.8\textwidth]{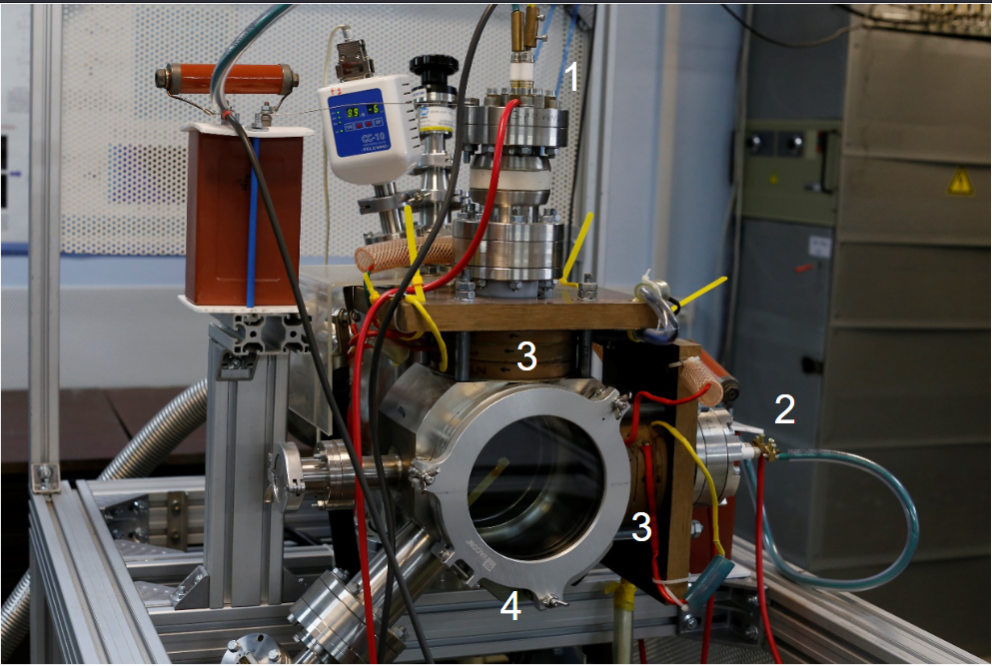}
\caption{An image of the experimental setup. 1, 2 -- plasma sources, 3 -- magnetic coils, 4 -- plasma vacuum chamber.\label{fig:setup}}
\end{figure}

The experiment is carried out in a vacuum chamber with the ambient pressure below $10^{-6}$ Torr. The experimental setup is shown in Fig. \ref{fig:setup}. The plasma is generated by a specially designed generator as a result of an arc discharge. The cylindrical generator has an outlet opening with a diameter of 2 cm. It is isolated from the chamber walls by a high-voltage insulator. The cathode, which is the plasma source, is made of aluminum. A detailed study of the plasma generator can be found in \cite{Viktorov2017}. In a typical operating mode, it is capable of generating a plasma flow with a velocity of $\approx 1.5\times 10^6$ cm/s and an electron temperature of about 3 eV. The measured average ion charge is $Z=1.7$. Since the energy of the directed ion motion is thus about $33$ eV, the flow is supersonic with a Mach number $M \approx 3.5$. By adjusting the discharge capacitor voltage from hundreds of volts to 5 kV, it is possible to change the discharge current value and, as a consequence, the ion concentration in the flow within the range of $10^{13}$--$10^{16}$ cm$^{-3}$. As a voltage changes, the flow velocity, though, remains practically constant. The typical discharge duration is 20 $\mu$s, which allows generating a plasma bunch about 30 cm long, which is comparable to the dimensions of a vacuum chamber, so that during the discharge the plasma is able to completely fill it.

The magnetic field is created by two pulsed coils of the same design and placed at the right angle to each other. The current pulse duration in them is about 3 ms. The magnetic field strength in the center of the coils reaches 3.3 T, but in the chamber it is 10--100 times smaller. For instance, for a discharge current of 1 kA, the field in the center of the chamber is 23 mT.

\begin{figure}
\includegraphics[width=0.8\textwidth]{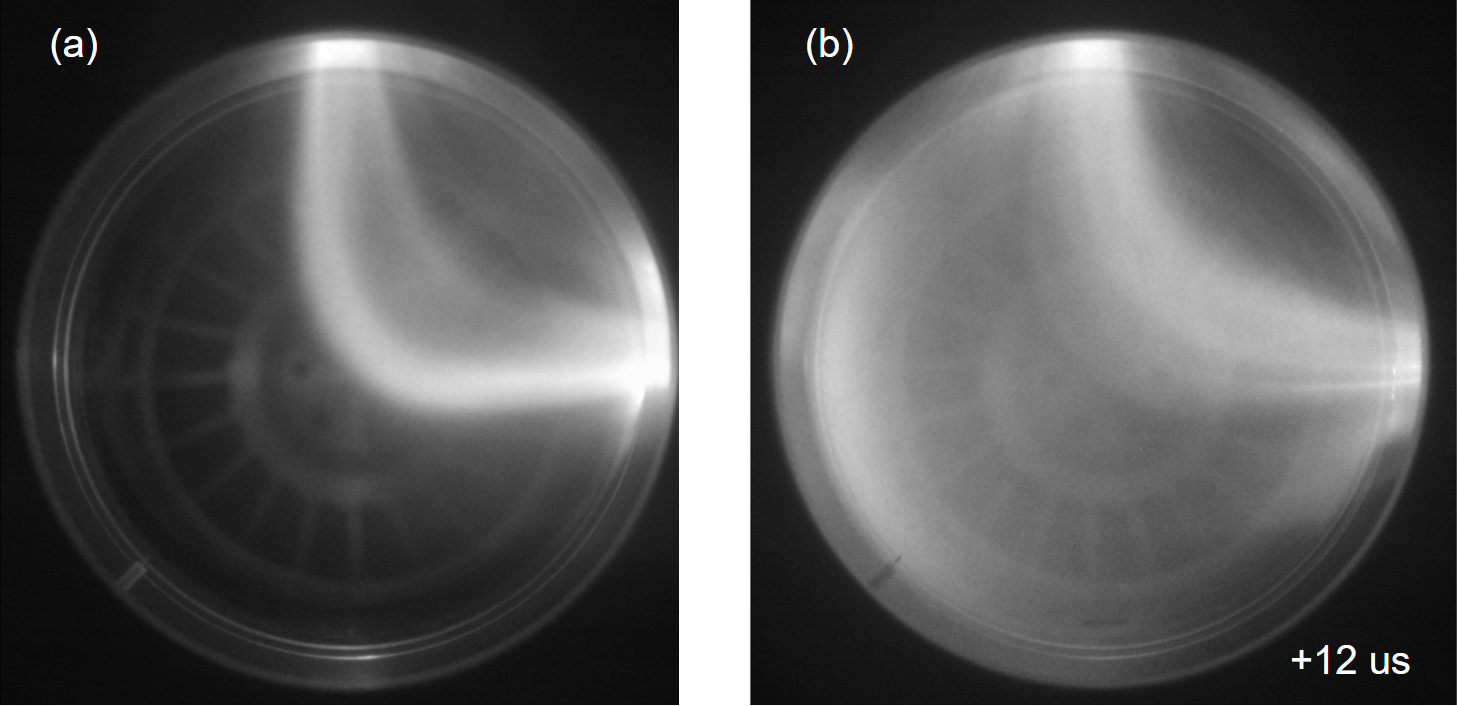}
\caption{Images of optical plasma glow during collision of two plasma flows at different moments in time. (a) Photograph at the moment of maximum arc current of the vacuum-arc discharge, (b) photograph at the moment of time later by 12 $\mu$s relative to (a). Exposure duration is 1 $\mu$s. In the image (b) the formation of plasma filaments in the form of bright luminous threads is clearly visible.\label{fig:discharge}}
\end{figure}

In the experiment, the arc discharge is initiated with some delay relative to the current discharge in the coils. By this time, a stationary magnetic field of an arched configuration is established in the chamber. In the studied configuration, two plasma generators are located inside the coils, and the plasma flows generated by them are injected from the bases of the arch along the magnetic field lines. In preliminary experiments, at low current in the magnetic coils, the chamber was filled with plasma, observed by its optical glow. At high current values in the coils and, consequently, for a strong magnetic field, a glow in the form of an arch is observed as shown at Fig. \ref{fig:discharge}. Thus, one of the goals of this work will be to study the transition between these two regimes.

\section{\label{sec:methods}Numerical methods}

Numerical modeling of weakly collisional highly non-equilibrium plasma generally requires consistent consideration of kinetic effects, which can be ensured by numerical modeling of Vlasov---Maxwell system of equations \cite{Bertrand2019}:
\begin{gather}
    \frac{\partial f_s}{\partial t} + \mathbf v_s\frac{\partial f_s}{\partial \mathbf r} + \frac{q_s}{m_s}\left(\mathbf E + \mathbf v_s\times\mathbf B\right)\frac{\partial f_s}{\partial \mathbf v_s} = 0 \\
    \frac{1}{c^2}\frac{\partial \mathbf E}{\partial t} = \mathrm{rot}\,\mathbf B - \mu_0\mathbf j \\
    \frac{\partial \mathbf B}{\partial t} = \mathrm{rot}\,\mathbf E \\
    \mathbf j = \sum_s q_s\int \mathbf v_sf_s(t,\mathbf r,\mathbf v_s)\mathrm d \mathbf v_s
\end{gather}
where index $s=e,i$ refers to electrons and ions respectively, $f_s(t,\mathbf r,\mathbf v_s)$ are particle distribution functions, $q_s, m_s, \mathbf v_s$ are electric charge, mass and velocity of particles, $\mathbf E(t,\mathbf r), \mathbf B(t,\mathbf r)$ are macroscopic electric and magnetic fields, $c,\mu_0$ are speed of light and magnetic constant, $\mathbf j(t,\mathbf r)$ is current density and integration is assumed to be over all velocities.

This system is usually treated numerically by means of particle-in-cell method \cite{Birdsall2004} or some kind of grid-based methods \cite{Palmroth2018}. However, performing full-scale modeling with realistic parameters using a fully kinetic electromagnetic code turns out to be technically challenging even in two-dimensional geometry due to relatively low velocities of plasmas: a typical plasma frequency of electrons is of the order of $10^{12}$ s, the corresponding time step should be less than 1 ps, i.e., modeling processes with a duration of about 10 $\mu$s requires tens of millions of iterations, which is thousand times greater than the typical capabilities of modern programs.

In this regard, within the framework of this study, it was decided to conduct a full-scale modeling using a hybrid method, in which only ions are described kinetically, and electrons are represented as a massless neutralizing liquid. To partially account for kinetic effects, however, electrons were described in a so-called 10-moment approximation taking into account the evolution of their pressure tensor. We also neglect displacement current $\partial \mathbf E/\partial t$ in Maxwell equations, assuming that all velocities are much less than speed of light and the system can be described in a so-called Darwin (low-frequency) approximation \cite{Krause2007}. The governing equations then become \cite{Hesse1995}:
\begin{gather}
    \frac{\partial f_i}{\partial t} + \mathbf v_i\frac{\partial f_i}{\partial \mathbf r} + \frac{Ze}{m_i}\left(\mathbf E + \mathbf v_i\times\mathbf B\right)\frac{\partial f_i}{\partial \mathbf v_i} = 0 \\
    n_e = Zn_i = Z\int f_i(t,\mathbf r,\mathbf v_i)\mathrm d \mathbf v_i \\
    \mathbf V_i = \frac{1}{n_i}\int \mathbf v_if_i(t,\mathbf r,\mathbf v_i)\mathrm d \mathbf v_i \\
    \mathbf E = -\mathbf V_i\times\mathbf B + \frac{1}{en_e}\left(\mathbf j\times\mathbf B - \nabla.\mathbb{P} \right) \\
    \frac{\partial \mathbf B}{\partial t} = \mathrm{rot}\,\mathbf E \\
    \mathbf j = \frac{1}{\mu_0}\mathrm{rot}\,\mathbf B \\
    \mathbf V_e = -\frac{1}{en_e} \mathbf j + \mathbf{V_i} \\
    \frac{\partial \mathbb{P}}{\partial t} + \mathbf V_e.\nabla\mathbb{P} = - \mathbb{P}\nabla.\mathbf V_e - \mathbb{P}.\nabla\mathbf V_e - \left(\mathbb{P}.\nabla\mathbf V_e\right)^T - \frac{e}{m_e}\left[\mathbb{P}\times\mathbf B + \left(\mathbb{P}\times\mathbf B\right)^T\right]
\end{gather}
where $n_{e,i}(t,\mathbf r)$ are electron and ion densities, $\mathbf V_{e,i}(t,\mathbf r)$ are averaged (hydrodynamical) electron and ion velocities, $\mathbb{P}$ is a pressure tensor of electrons, $\nabla$ is a del (nabla) vector differential operator and $(\cdot)^T$ means tensor transposition. We would like to underline that despite displacement current is neglected it is not a purely electrostatic approximation as we still have time-dependent electric field and inductive magnetic field. It allows to model low-frequency magneto-hydrodynamics waves and instabilities as well as Weibel-like instabilities with somewhat corrected dispersion relations. Those corrections, however, are of the order of $V_{e,i}/c$ \cite{Sarrat2016,Sladkov2023}.

The two-dimensional hybrid simulation was carried out by the AKA code \cite{Sladkov2020}. This code was previously successfully tested on a model problem for analyzing the magnetic reconnection process in the perturbed Harris sheet on electron spatial scales \cite{Sladkov2021}, in the problem of long-term dynamics of the Weibel instability \cite{Sladkov2023}, as well as in modeling experiments on laser ablation of plasma in an external magnetic field \cite{Bolanos2022,Burdonov2022,Zemskov2024} and in self-generated magnetic fields \cite{Sladkov2024}.

The hybrid simulations were close to the conditions of the real experiment. The simulation domain size corresponded to a region of $50\times 50$ cm, the initial ion concentration in the plasma flows was $10^{16}$ cm$^{-3}$, the plasma was assumed to be fully ionized and consist of singly ionized aluminum. Plasma flows were generated at the boundary of the simulation domain. The initial plasma velocity was varied in different simulations near the value of $10^{6}$ cm/s, the temperature of electrons and ions was assumed to be initially zero. The total simulation time was about 80 $\mu$s. The external magnetic field was considered to be given, its configuration was obtained by means of a separate electromagnetic calculation for the known geometry of real coils. Its value in the simulation domain was about 0.25 T, which corresponds to the estimate of the magnetic field in the center of the chamber in the experiment. The boundaries were supposed to be open for particles and fields.

For a qualitative examination of electron kinetic effects, a fully kinetic electromagnetic simulation was also carried out by the particle-in-cell method \cite{Birdsall2004}. This required scaling the system to reduce the computation time. Two main approximations were made: first, the flow velocity was increased to bring it closer to the speed of light, while remaining nonrelativistic – in the simulation it was equal to $10^{9}$ cm/s; and second, the ion-electron mass ratio was reduced to 16, which allowed the ion and electron scales to be brought closer to each other. The remaining quantities were chosen so as to preserve the main dimensionless parameters without significant changes: the ratio of the magnetic Mach number is on the order of unity, the ratios of the ion gyroradius and ion inertial length to the arch width are on the order of unity, and the ratio of the electron gyroradius to the arch width is less than unity. The detailed comparison of dimensional and dimensionless quantities in hybrid and kinetic simulations are given in Table \ref{tab:comparison}.

\begin{table*}
\caption{\label{}Comparison of parameters of hybrid and kinetic simulations performed in the study\label{tab:comparison}}
\begin{ruledtabular}
\begin{tabular}{lcc}
\textbf{Parameter}	& \textbf{Hybrid}	& \textbf{Kinetic}\\
\hline
$m_i$		& 26.98 a.u.m.			& 26.98 a.u.m.\\
$m_i/m_e$		& 100			& 16\\
Initial flow velocity, $V_0$		& $10^6$ cm/s\footnote{The flow velocity in hybrid simulations was varied. Here the typical value is given.}			& $10^9$ cm/s\\
Initial particle density, $N_0$		& $10^{16}$ cm$^{-3}$	& $10^{16}$ cm$^{-3}$\\
Initial plasma beam width, $w_0$	& 7.5 cm			& 0.30 cm\\
Magnetic field in injection point, $B_0$	& 0.25 T			& 250 T\\
Simulation domain, L	& 47.4 cm			& 2.84 cm\\
Alfven velocity, $V_A = B_0/\sqrt{\mu_0m_iN_0}$	& $1.05\times 10^6$ cm/s & $1.05\times 10^9$ cm/s\\
Ion inertial length, $d_0 = c\sqrt{\varepsilon_0m_i/e^2N_0}$\footnote{Here $\varepsilon_0$ is an electric constant.}	& 1.18 cm			& 1.18 cm\\
Ion gyroradius, $r_i = m_iV_0/eB_0$	& 1.12 cm			& 1.12 cm\\
Electron gyroradius, $r_e = m_eV_0/eB_0$	& 1.12 mm			& 0.07 cm\\
Magnetic Mach number, $\mathrm{M_m} = V_0/V_A$     	& 0.95			& 0.95\\
$V_0/c$	                        & $3.34\times 10^{-5}$			& $3.34\times 10^{-2}$\\
$w_0/d_0$	                    & 6.36			& 0.25\\
$w_0/r_i$	                    & 6.70			& 0.27\\
$w_0/r_e$	                    & 670			& 4.29\\
\end{tabular}
\end{ruledtabular}
\end{table*}

The fully kinetic simulations were performed using the Smilei code \cite{Smilei}, which implements a fully electrodynamic, relativistic particle-in-cell method. The code allows to inject particles during the calculation, but there is no way to specify an external constant field. Therefore, the external magnetic field was simulated by a system of superheavy charged particle currents. The size of the simulation region in both coordinates was 2.4 ion inertial lengths = 9.6 electron inertial lengths. The temperature of the injected particles was zero. The boundary conditions were open for both particles and fields.

\section{\label{sec:results}Simulations results}

A series of calculations were performed using a hybrid numerical modeling method, in which the number of flows and their speed were varied. Note that for the selected parameters of plasma flows, the pressure of an individual flow and the magnetic field pressure in the center of the interaction region are comparable. Accordingly, the global nature of the interaction depends on whether the flow pressure exceeds the magnetic pressure or not. Two typical regimes were observed in the modeling: subcritical at lower flow pressure and overcritical at higher flow pressure.

\subsection{\label{sec:subcritical}Subcritical interaction regime}

\begin{figure}
\includegraphics[width=0.7\textwidth]{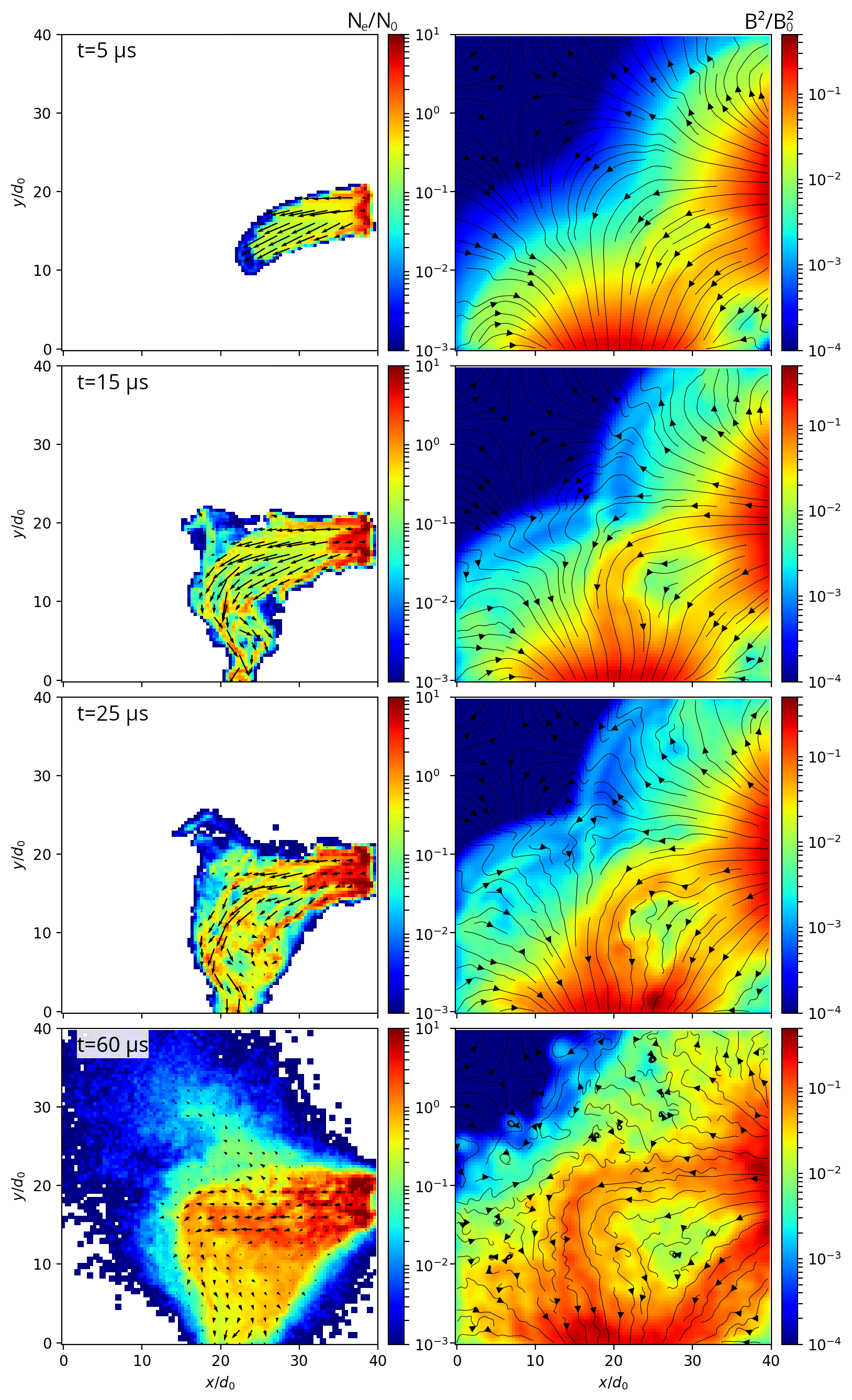}
\caption{\label{fig:one-flow}Plasma particle concentration (left) and magnetic pressure (right) at different time instants for one flow. The arrows on the left show the vector of the average ion velocity, on the right – the magnetic field lines. The concentration is normalized to $N_0 = 10^{16}$ cm$^{-3}$, the square of magnetic field is normalized to $B_0^2 = (250\ \mathrm{mT})^2$, the spatial coordinates are normalized to $d_0=1.2$ cm.}
\end{figure}

Fig. \ref{fig:one-flow} shows the calculation result for one flow with a velocity of $10^{6}$ cm/s. It is evident that the flow as a whole follows the magnetic field lines. At the same time, the plasma concentration during its expansion decreases, since the magnetic field in the center of the calculation region is lower than at the plasma ejection point. The plasma expansion is accompanied by a compression of the magnetic field at both boundaries of the plasma tube being formed. However, at its apex, at late times, a partial plasma ejection is observed.

\begin{figure}
\includegraphics[width=0.8\textwidth]{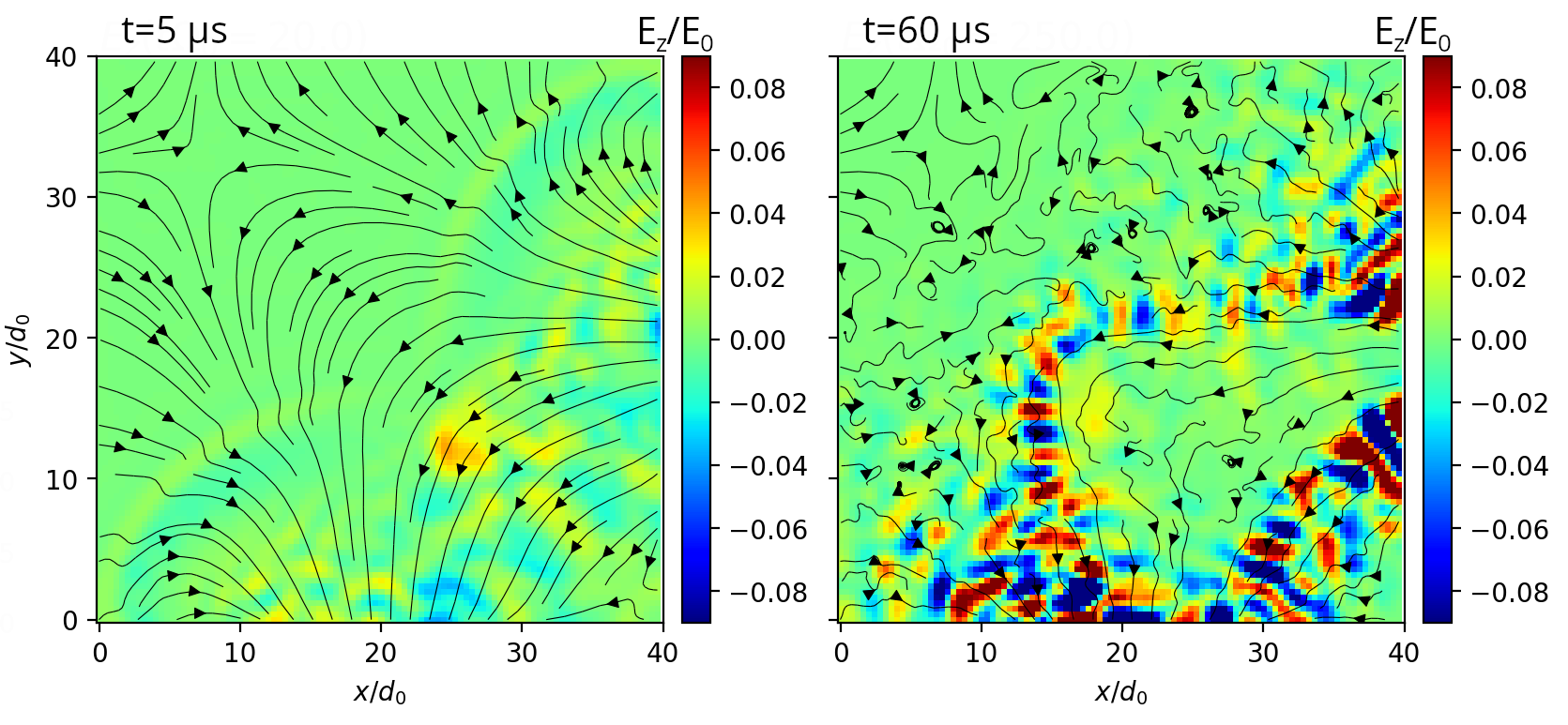}
\caption{\label{fig:flow-polarize}$z$-component of electric field at times of 5 $\mu$s (left) and 60 $\mu$s (right). Arrows show the magnetic field lines. Electric field is normalized to $E_0 = V_0B_0 = 25$ V/cm.}
\end{figure}

The transverse plasma escape is explained by the $E\times B$ drift, which occurs as a result of inductive effect: as the plasma tends to push magnetic fields aside due to diamagnetic effect, the time-dependent magnetic field induces transversal electric field $E_z$. This field is clearly visible in Fig. \ref{fig:flow-polarize} at the time of 5 $\mu$s.

In the same figure at later times it is seen that a strong alternating field arises at the arch bottom boundary, polarized perpendicular to the plane and propagating along the magnetic field. This wave has a wavelength close to the ion inertial length $d_i = c\sqrt{\varepsilon_0 m_i/e^2n_i}$, and therefore it can be seen as a surface low-frequency mode excited at frequency close to ion cyclotron frequency $\omega_{ci} = eB/m_i$. Indeed in the limit of zero frequency for Alfven waves we have $\omega = kV_A$, where $\omega$ is a wave frequency, $k$ is a wave number and $V_A = B/\sqrt{\mu_0 m_i n_i}$ is Alfven velocity. For $k=1/d_i$ we get $\omega = \omega_{ci}$. This ion cyclotron wave is expected to grow in the plasma with the anisotropic ion distribution function. Particularly, it should be unstable if $T_\parallel > T_\perp$ ($T_\parallel$ and $T_\perp$ being dispersions of particle velocities along and perpendicular to local magnetic field) \cite{Sagdeev1961}. Here due to short times and low loss rate of particles through arc ends we have for ions $T_\parallel > T_\perp$ as they are injected along magnetic lines with zero transverse momentum so it is indeed unstable and can grow.

\begin{figure}
\includegraphics[width=0.8\textwidth]{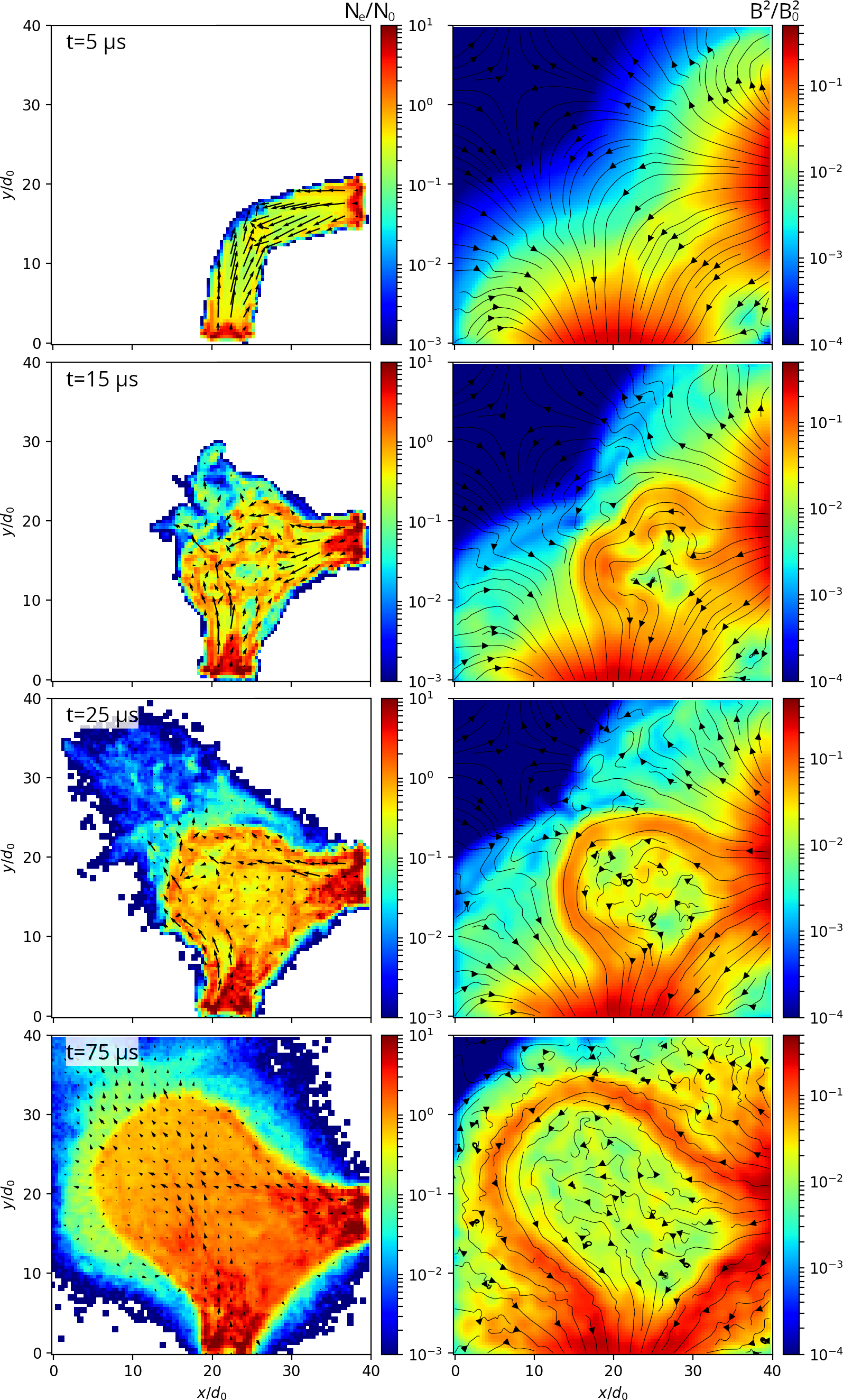}
\caption{\label{fig:low-pressure-flows}The same as in Fig. \ref{fig:one-flow} for the case of two flows.}
\end{figure}

Fig. \ref{fig:low-pressure-flows} shows the simulation results for two counter-propagating flows with the same interaction parameters. Note that for the first 15-20 $\mu$s the dynamics with two flows is almost the same as the dynamics with one flow. A plasma-filled tube is formed, at the boundaries of which the magnetic field is compressed by the plasma pressure. However, at later times the expansion of the outer side of the tube occurs much faster than in the case of one flow, which is due to the doubled pressure of the plasma flows.

\begin{figure}
\includegraphics[width=0.8\textwidth]{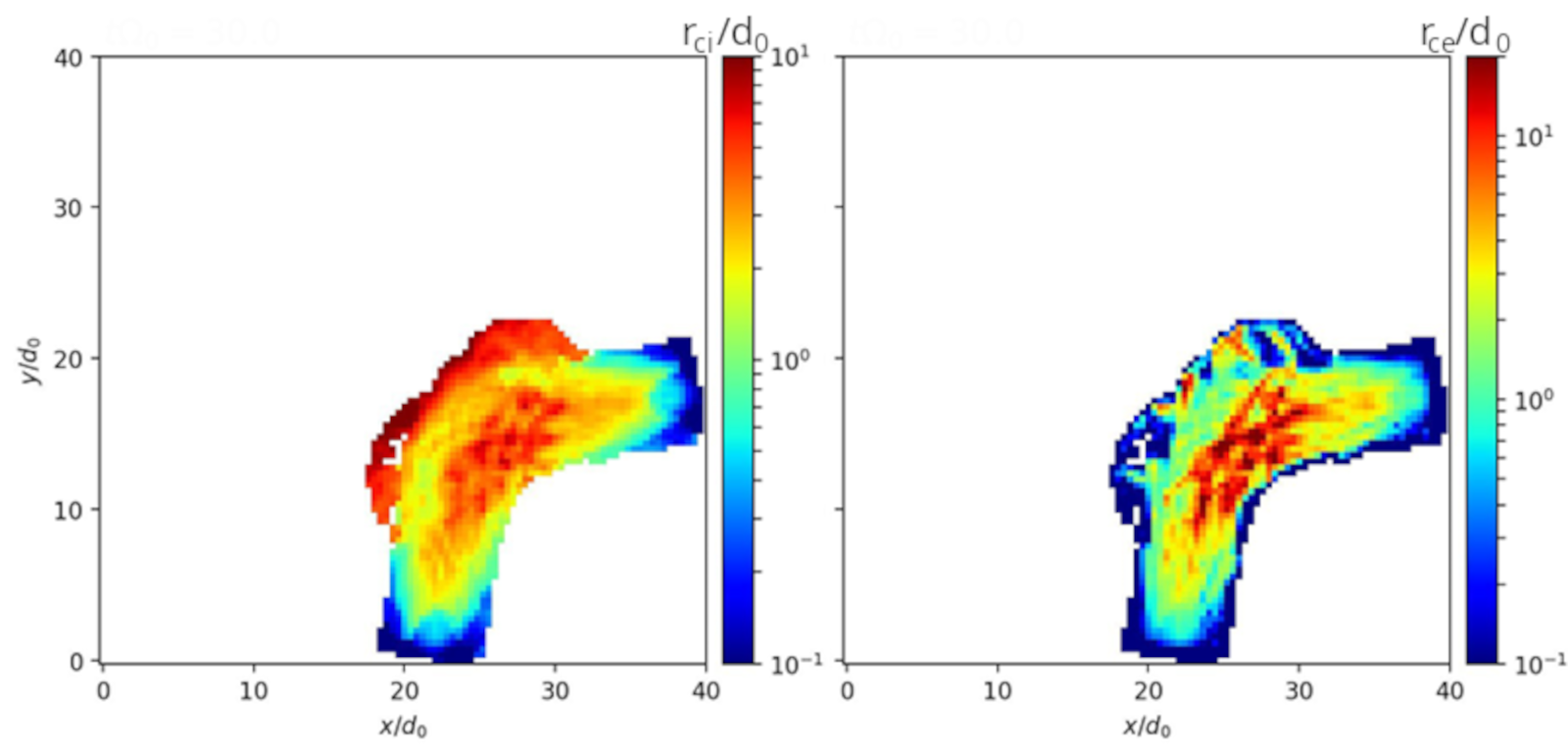}
\caption{\label{fig:low-pressure-flows-2}Local gyroradius of ions (left) and electrons (right) normalized to ion inertial length $d_0=1.2$ cm for the case of a collision of two flows at 7.5 $\mu$s after the start of the simulation. The parameters are the same as in Fig. \ref{fig:low-pressure-flows}.}
\end{figure}

As can be seen from Fig. \ref{fig:low-pressure-flows-2}, the ions in such a configuration are demagnetized, since their gyroradius is comparable to the size of the tube, and the electrons are effectively magnetized -- the plasma beta, although large in the center of the tube, is close to unity at its boundary, where the field is compressed.

At later times, the tube is stretched and forms a configuration with oppositely directed magnetic fields. The zero-magnetic-field line is formed along diagonal of the simulation box. Magnetic reconnection, however, is slow at these parameters. The tearing instability is not observed, so the formation of plasmoids and the separation of the top of the tube from the base do not occur, the configuration slowly evolves, but remains stable even 70 $\mu$s after the onset of interaction. Such stability is associated with a relatively low pressure of the plasma flows.

\subsection{\label{sec:overcritical}Overcritical interaction regime}

\begin{figure}
\includegraphics[width=0.8\textwidth]{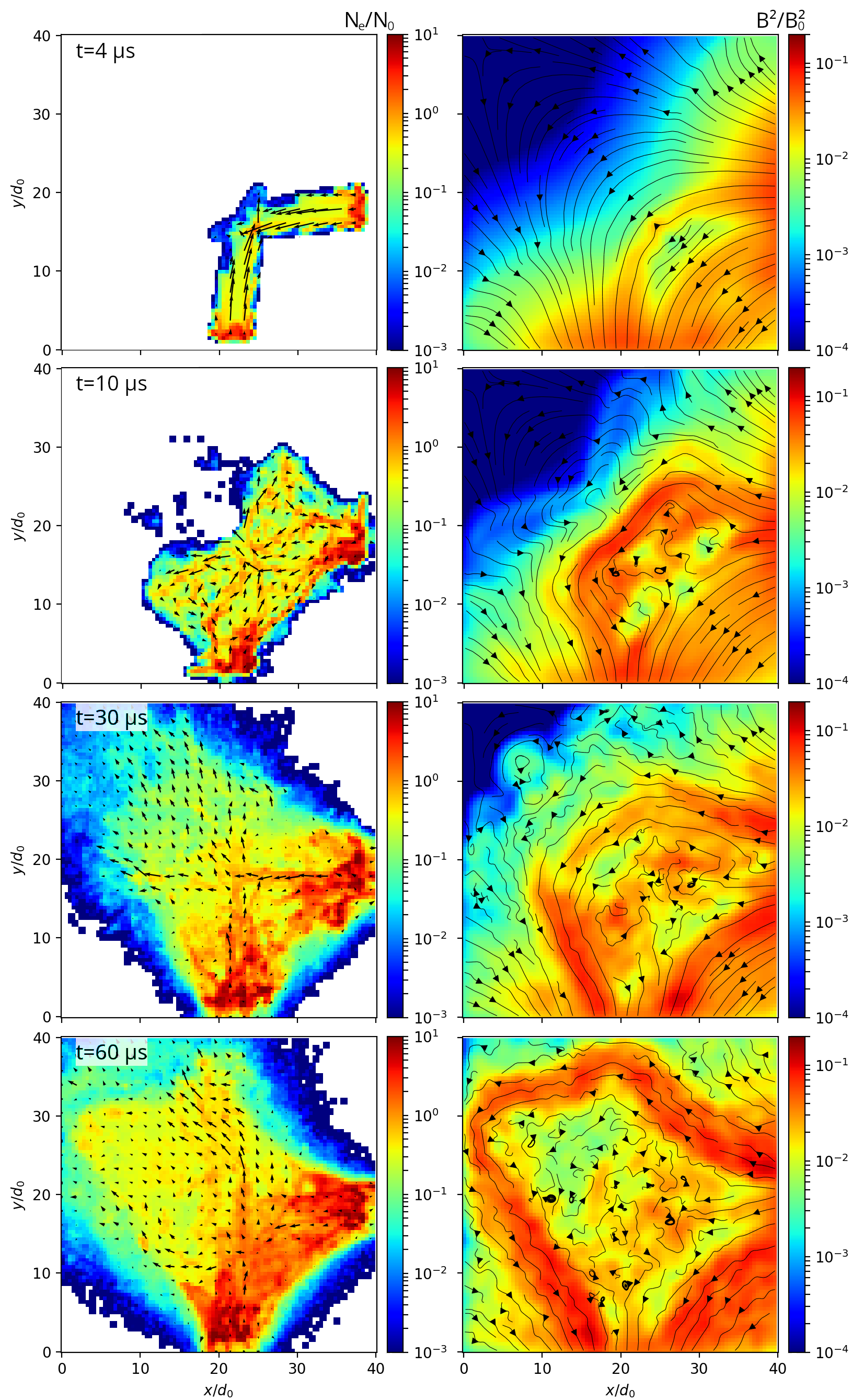}
\caption{\label{fig:high-pressure-flows}Same as in Fig. \ref{fig:low-pressure-flows} for flows with twice the speed.}
\end{figure}

Fig. \ref{fig:high-pressure-flows} shows the simulation result for a plasma flow velocity doubled and, accordingly, four times higher pressures of these flows. It is evident that in this case, a quasi-stationary tube is not formed. Almost immediately, the process of separation of a part of the flows with the formation of plasmoids begins. At later times, a configuration with an elongated tube is formed, but its shape is deformed, and the magnetic field pressure inside is higher. Clearly visible plasmoids with a closed magnetic field configuration break away from the top of the tube (see the upper left corner at time instant 30 $\mu$s) which is a manifestation of the tearing instability happened in reverse magnetic field lines configuration. Inside the plasma we see an intense mixing of magnetic field lines with a quasi-turbulent state arising. The hybrid code, though, does not allow to investigate the turbulization as it does not resolve electron spatial scales.

\begin{figure}
\includegraphics[width=0.8\textwidth]{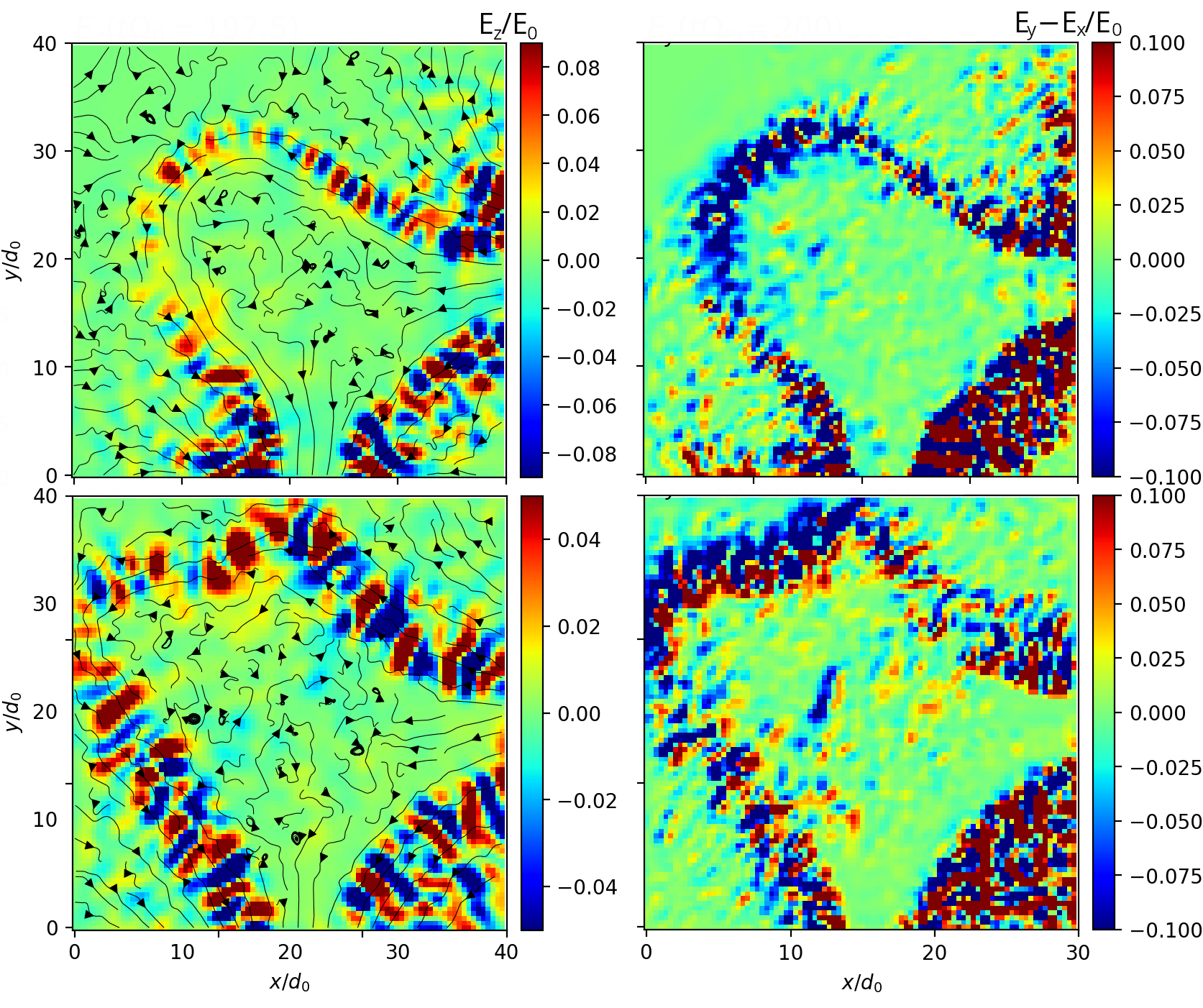}
\caption{\label{fig:alfven-waves}$z$-component (left) and in-plane component (right) of the electric field at time 50 $\mu$s after the start of the calculation for two flows. Top row – for the parameters of Fig. \ref{fig:low-pressure-flows}. Bottom row – for parameters of Fig. \ref{fig:high-pressure-flows}. Electric fields are normalized to $E_0 = 25$ V/cm.}
\end{figure}

As in the case of one flow, the formation of a plasma tube in the case of two flows is accompanied by resonant excitation of surface waves at ion cyclotron frequency. This can be seen in Fig. \ref{fig:alfven-waves}. Note that the excitation of these waves is observed for both relatively slow flows and faster ones. There the in-plane component of the electric field is also shown. It is also alternating at tube boundaries which is expected as the waves at this range of frequencies should be elliptically polarized. We also note that the in-plane component exceeds the $z$-component which means that this is a mode connected to Alfven wave at lower frequencies. In other words this is an ordinary elliptically polarized wave.

\subsection{\label{sec:weibel}Development of Weibel instability}

A feature of the two-flow interaction compared to the case of one flow is a much more pronounced density filamentation along the flows (cf. Fig. \ref{fig:one-flow} and Fig. \ref{fig:low-pressure-flows} at 15 $\mu$s). We attribute it to Weibel instability. In the simulation performed, electrons are described hydrodynamically, but taking into account the second-order moments - the components of the pressure tensor. In this case, the Weibel instability manifests itself in the presence of anisotropy of this tensor.

\begin{figure}
\includegraphics[width=0.8\textwidth]{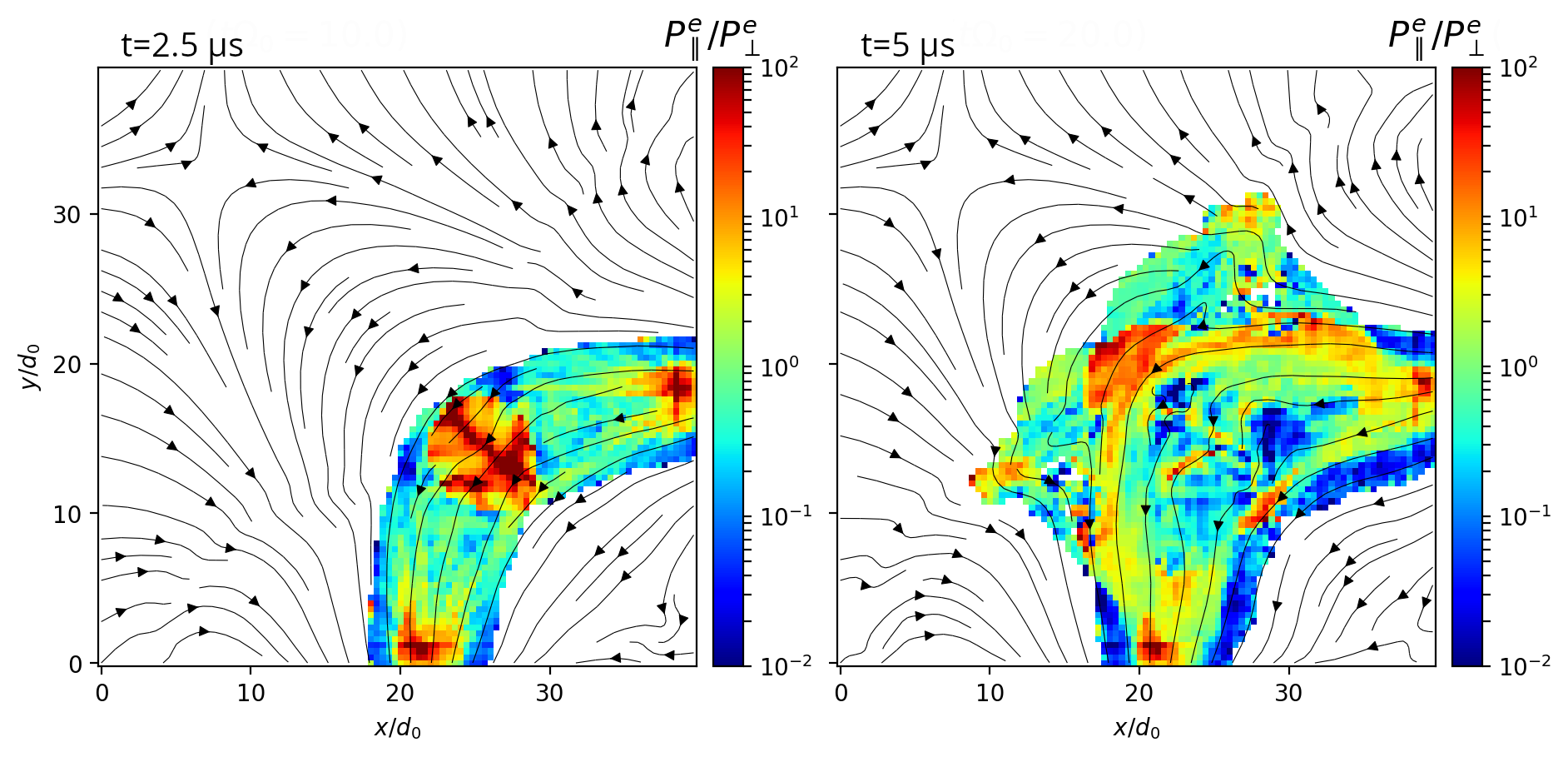}
\caption{\label{fig:anisotropy}Ratio of the longitudinal component of the pressure tensor with respect to the magnetic field to the transverse component at two time instants.}
\end{figure}

Fig. \ref{fig:anisotropy} shows the distribution of the anisotropy of the pressure tensor obtained during the simulation. It is evident that a highly anisotropic plasma is formed when the flows collide. At later times, however, the anisotropy in this region of space drops to unity, indicating that Weibel instability has developed and reached saturation by this time. Initially, the instability leads to filamentation of currents, which then give rise to filamentation of density.

Weibel instability, however, in these conditions cannot be correctly described in hydrodynamic approximation for electrons, since their distribution function, at least in the early stages of the development of instability, is far from equilibrium. For a qualitative consideration of the kinetically determined features of the developed Weibel instability, fully kinetic modeling was carried out by the particle-in-cell method in the model conditions described in the section \ref{sec:methods}.

\begin{figure}
\includegraphics[width=0.8\textwidth]{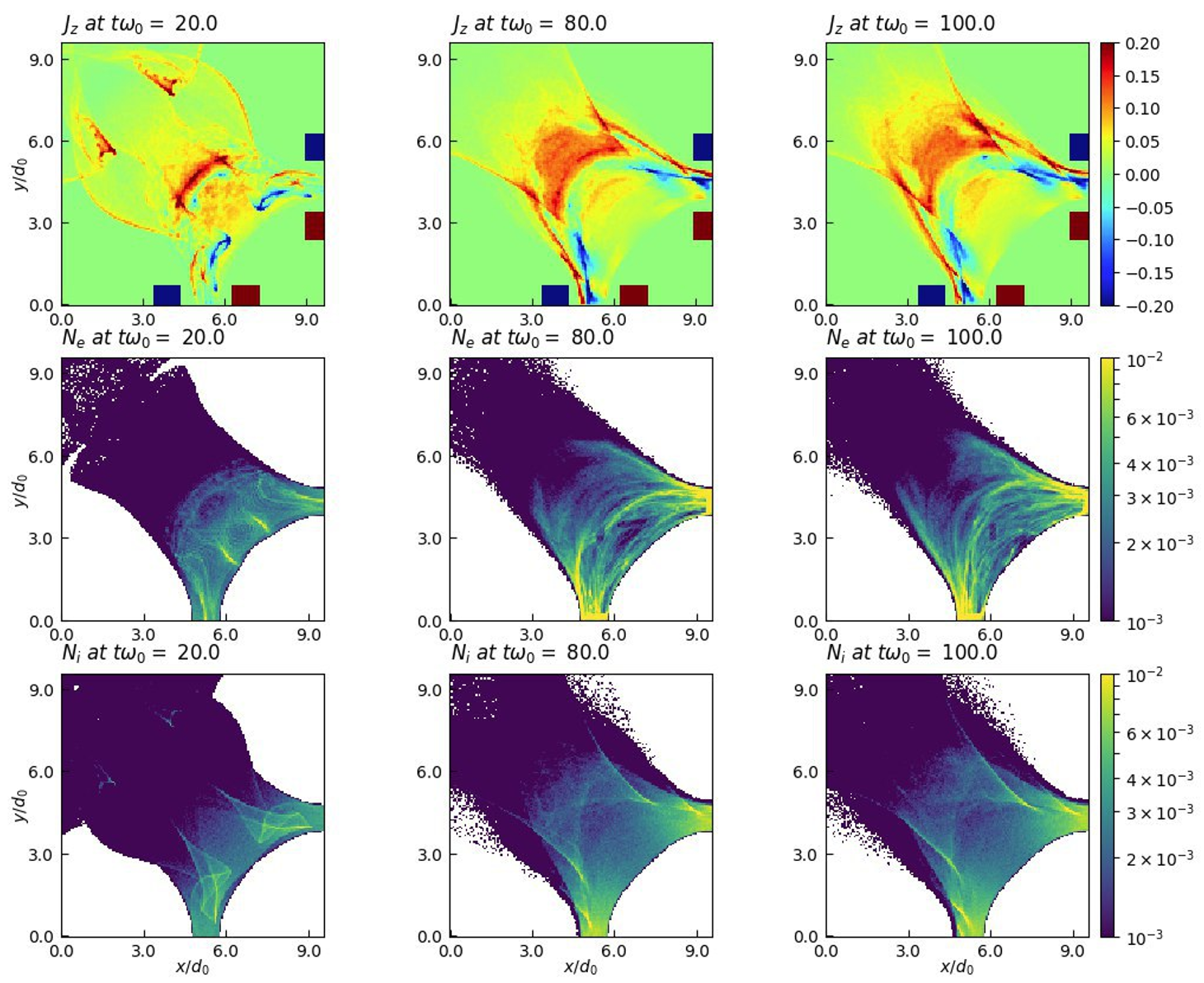}
\caption{\label{fig:pic}The result of fully kinetic modeling. $z$-component of current density (top row), the concentration of electrons (middle row) and ions (bottom row) at three time instants are shown. The current density $J_z$ is normalized to $eN_0V_0$ ($N_0$, $V_0$ are initial particle concentration anf velocity of flows), the concentrations $N_e$, $N_i$ are normalized to $N_0$, time $t$ is normalized to inverse electron plasma frequency $\omega_0^{-1} = \sqrt{\varepsilon_0m_e/e^2N_0}$ and spatial coordinates $x$, $y$ are normalized to electron inertial length $d_0=c/\omega_0$.}
\end{figure}

Kinetic simulation results are shown in Fig. \ref{fig:pic}. It can be seen that the structure of the tube as a whole coincides with the results of hybrid modeling. At the same time, filamentation of electronic density is much stronger, and the width of the filaments is generally smaller. This is due to a more pronounced anisotropy of the distribution function in the case of a kinetic description of the plasma.

It should be noted that in our setup Weibel instability is developing in the presence of external magnetic field. In this case the growth of modes with wavelengths lower than gyroradius is inhibited so it is expected that the mode with a wavenumber $k=1/r_{ce}$ will be the most pronounced. This is in coincidence with the filament width observed in the simulation.

\begin{figure}
\includegraphics[width=0.8\textwidth]{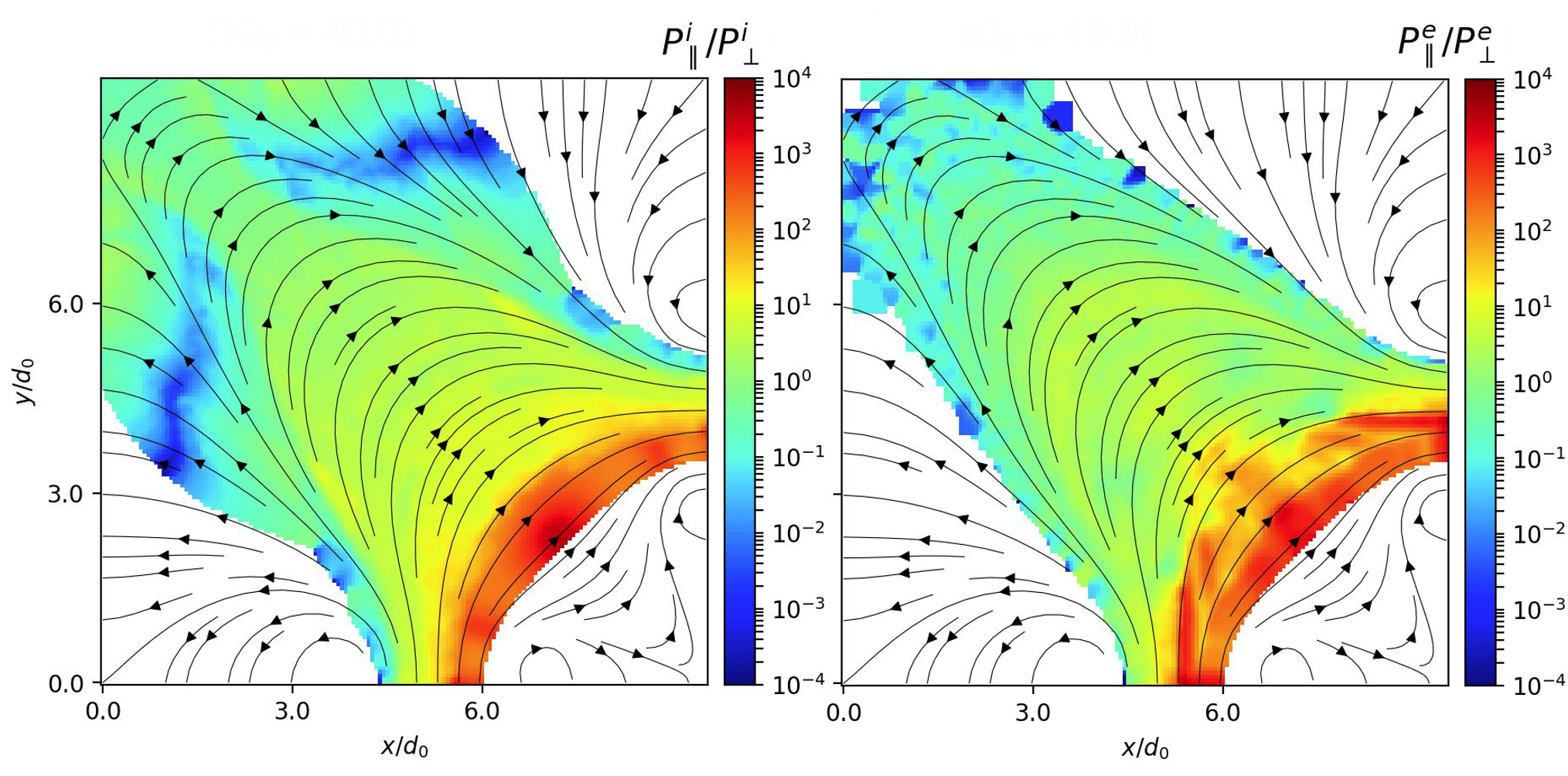}
\caption{\label{fig:pic-anisotropy}Ratio of the longitudinal (with respect to the magnetic field) component of the ion (left) and electron (right) pressure tensors to the transverse component in the calculation plane. The time instant $t=40\omega_0^{-1}$ after the start of the calculation is shown (see Fig.~\ref{fig:pic} for normalizations).}
\end{figure}

From the anisotropy distribution in Fig. \ref{fig:pic-anisotropy} it is evident that a plasma strongly anisotropic along the field is formed in the tube. Electrons and ions circulate along the tube, reflecting from its ends. As a result of this movement, their flow velocity transforms into a thermal velocity longitudinal to the magnetic field, which remains high for a long time relative to the transverse velocity due to the absence of collisions.

\begin{figure}
\includegraphics[width=0.8\textwidth]{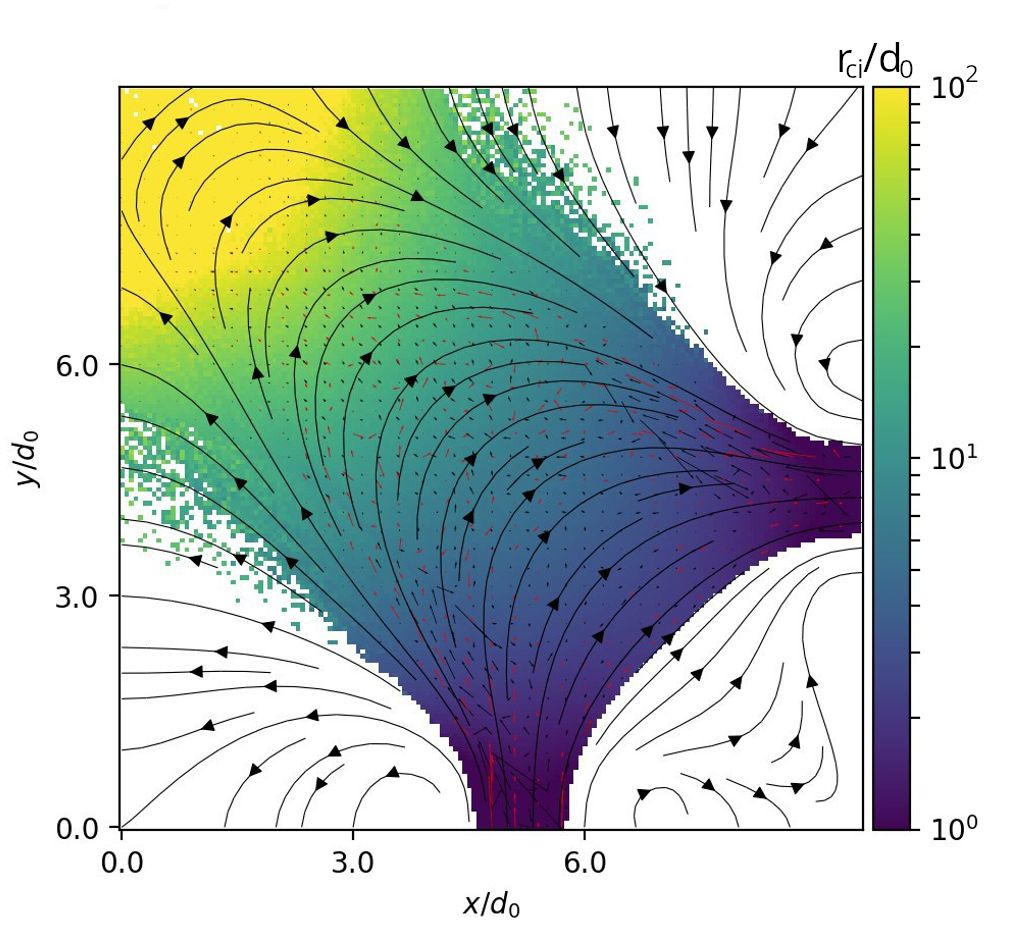}
\caption{\label{fig:pic-larmor}Local ion gyroradius at the time instant $t=100\omega_0^{-1}$ after the start of the calculation (see Fig.~\ref{fig:pic} for normalizations).}
\end{figure}

Note that here we also observe a slow plasma outflow from the top of the tube and it is mainly associated with the large gyroradius of the ions. In Fig. \ref{fig:pic-larmor} it is seen that it is of the order of the transverse size of the tube, and accordingly, along with the electric drift, a fairly strong gradient drift is observed.

\section{\label{sec:discussion}Discussion}

Let us summarize the main conclusions obtained in the modeling. In the experiment, two interaction regimes should be observed: subcritical, corresponding to a magnetic Mach number $M_m < 1$, at which the plasma flow pressure is less than the magnetic pressure, and overcritical, corresponding to $M_m > 1$, at which the plasma flow pressure is higher than the magnetic pressure. In the subcritical regime, the magnetic arch is filled with plasma with the formation of a plasma arch, and in the overcritical mode, the plasma breaks through the magnetic lines and fills the entire volume of the chamber. Note that these regimes have already been observed in preliminary experiments with a discharge in magnetic fields of different magnitudes.

In the subcritical regime, due to the diamagnetic effect, the magnetic field at the boundary of the arch is compressed, which is also accompanied by the formation of denser plasma regions. In the experiment, such thickenings can be observed as a brighter glow of the edges of the arch as seen at Fig. \ref{fig:discharge}. Along the boundaries of the arch, excitation of a wave near the ion cyclotron frequency is observed, which can likely be recorded in the experiment by a MHz-range receiver. The top of the plasma arch slowly moves outward due to the $E\times B$ drift, which can be recorded in the experiment by a high-speed camera with a submicrosecond shutter, filming the plasma with a variable delay after the start of the discharge. At low values of the magnetic Mach number, however, the speed of this drift will be small due to the relative weakness of the diamagnetic effect.

The plasma arch demonstrates the most interesting dynamics in the near-critical mode. In this case, the arch expansion rate is comparable to the initial flow velocity and the Alfven velocity. During the discharge, a significant expansion of the plasma is observed with the formation of a region with oppositely directed magnetic fields. At magnetic Mach numbers slightly greater than unity, intense turbulence is observed in this region, accompanied, among other things, by the formation of plasmoids breaking away from the arch. It can be expected that in this case, intense high-frequency radiation will be observed at frequencies near the electron cyclotron resonance, which can be recorded by a GHz-range receiver. In addition, it can be expected that high-energy electrons will be generated, which can also be recorded by an electron spectrometer.

In the plasma tube, the development of Weibel instability with the formation of plasma filaments is also expected. These filaments are also visible in experiment at Fig. \ref{fig:discharge} by their enhanced optical glow.

\section{Conclusions}

A two-dimensional numerical simulation of the process of propagation of one and two counter-streaming flows in an external magnetic field with an arched configuration similar to the experimental one was carried out. Full-scale simulation was carried out using a hybrid method, in which ions are described kinetically, and electrons are described hydrodynamically taking into account the evolution of their pressure tensor. Electron kinetic effects were also investigated in a fully kinetic simulations using the particle-in-cell method in a scaled geometry with a reduced electron-to-ion mass ratio, increased flow velocities and magnetic field strength, and reduced scales of the interaction region. It was shown that as a result of plasma propagation, an arched plasma tube is formed, in which the magnetic field is compressed at the edges. The plasma propagating along the tube is polarized, resulting in an $E\times B$ drift outward from the tube. At low flow velocities, this leads to a slow expansion of the tube and its quasi-stationary evolution. As a result, a region with oppositely directed magnetic field lines is formed, but magnetic reconnection in such a configuration is slow. At higher flow velocities, however, the tube evolution is much faster, stationarity is not achieved, and magnetic reconnection is more intense. The development of tearing instability and the corresponding formation of plasmoids detached from the base of the arch is observed.

In both cases, generation of elliptically polarized surface waves along tube boundaries is observed at a frequency close to the ion cyclotron frequency. Inside the tube, density filamentation is observed, caused by the development of Weibel instability, due to the occurrence of electron pressure anisotropy during the interpenetration of plasma flows. Fully kinetic modeling at scaled parameters confirmed the main conclusions of the hybrid modeling.

\begin{acknowledgments}
This research was funded by Russian Science Foundation grant number 23-12-00317. The simulations were performed on resources provided by
the Joint Supercomputer Center of the Russian Academy of Sciences.
\end{acknowledgments}

\bibliography{arc}

\end{document}